# Lessons Learned from the Real-world Deployment of a Connected Vehicle Testbed

**Mashrur Chowdhury, Ph.D., P.E., F. ASCE**
Eugene Douglas Mays Endowed Professor of Transportation and
Professor of Civil Engineering, Professor of Automotive Engineering and
Professor of Computer Science
Glenn Department of Civil Engineering, Clemson University
216 Lowry Hall, Clemson, South Carolina 29634
Tel: (864) 656-3313, Fax: (864) 656-2670; Email: mac@clemson.edu

**Mizanur Rahman\***
Ph.D. Candidate
Glenn Department of Civil Engineering, Clemson University
351 Flour Daniel, Clemson, SC 29634
Tel: (864) 650-2926; Fax: (864) 656-2670; Email: mdr@clemson.edu

**Anjan Rayamajhi, Ph.D. Student**
Department of Electrical and Computer Engineering,
314 Flour Daniel, Clemson University, SC 29634
Tel: (864)-633-8659, Fax: (864)-656-0145
Email: arayama@g.clemson.edu

**Sakib Khan**
Ph.D. Student
Glenn Department of Civil Engineering, Clemson University
351 Flour Daniel, Clemson, SC 29634
Tel: (864) 569-1082; Fax: (864) 656-2670; Email: sakibk@clemson.edu

**Mhafuzul Islam**
Ph.D. Student
Glenn Department of Civil Engineering, Clemson University
351 Flour Daniel, Clemson, SC 29634
Tel: (864) 986-5446; Fax: (864) 656-2670; Email: mdmhafi@clemson.edu

**Zadid Khan**
Ph.D. Student
Glenn Department of Civil Engineering, Clemson University
351 Flour Daniel, Clemson, SC 29634
Tel: (864) 359-7276; Fax: (864) 656-2670; Email: mdzadik@clemson.edu

**James Martin, Ph.D.**
Associate Professor
School of Computing, Clemson University,
211 McAdams Hall, Clemson, SC 29634, United States
Tel: 864-656-4529, Fax: +1-(864)-656-0145; Email: jmarty@clemson.edu

**ABSTRACT**

The connected vehicle (CV) system promises unprecedented safety, mobility, environmental, economic and social benefits, which can be unlocked using the enormous amount of data shared between vehicles and infrastructure (e.g., traffic signals, centers). Real-world CV deployments including pilot deployments help solve technical issues and observe potential benefits, both of which support the broader adoption of the CV system. This study focused on the Clemson University – Connected Vehicle Testbed (CU-CVT) with the goal of sharing the lessons learned from the CU-CVT deployment. The motivation of this study was to enhance early CV deployments with the objective of depicting the lessons learned from the CU-CVT testbed, which includes unique features to support multiple CV applications running simultaneously. The lessons learned in the CU-CVT testbed are described at three different levels: i) the development of system architecture and prototyping in a controlled environment, ii) the deployment of the CU-CVT testbed, and iii) the validation of the CV application experiments in the CU-CVT. Our field experiments with a CV application validated the functionalities needed for running multiple diverse CV applications simultaneously under heterogeneous wireless networking, and real-time and non-real-time data analytics requirements. The unique deployment experiences, related to heterogeneous wireless networks, real-time data aggregation, a distribution using broker system and data archiving with big data management tools, gained from the CU-CVT testbed, could be used to advance CV research and guide public and private agencies for the deployment of CVs in the real world.

**1.0 INTRODUCTION**

Connected vehicle technologies (CVT) offer unprecedented safety, mobility, environmental, economic and social benefits, which can be unlocked using the enormous amount of data shared between vehicles and infrastructure (e.g., traffic signals, centers) *(1)*. The real-world connected vehicle (CV) pilot deployments will materialize these promised benefits of the CVT, which are needed to make informed investment and deployment decisions for large-scale deployments by state or local public agencies. The United States Department of Transportation (USDOT) announced a proposed law to install vehicle-to-vehicle (V2V) and vehicle-to-infrastructure (V2I) communications using Dedicated Short Range Communication (DSRC) for all new lightweight vehicles to support different CV applications (i.e., safety, mobility and environmental applications) *(2)*. From the perspective of this federal agency, state transportation agencies should deploy CVT and use the CV-generated data from the deployment sites to support state-wide planning and policies *(3)*.

Although the USDOT has recently funded three pilot deployment sites in New York, Wyoming and Florida to evaluate the efficacy of CV applications *(4)*, the implementation guidelines for state or local agencies are still in the nascent stages of development. Initiated in Ann Arbor, Michigan in 2010, the CV Safety Pilot Program involved an analysis of the safety benefits of 2800 CVs at 29 infrastructure sites *(5)*. Although the deployment program was successful in terms of elucidating the safety benefits of CVs, this study does not provide detailed guidance for the cases where multiple CV applications must be operated simultaneously that would require heterogeneous wireless networking strategy for selecting specific communication options based on the corresponding application requirements in terms of data delivery delay and coverage range, data analytics capabilities for real-time data streaming, and data archiving strategies using big data management tools and infrastructure. Such deployment cases are required so that state public agencies may capitalize upon the benefits from multiple CV applications, and provide academic and research institutions the means to develop innovative CV applications without direct involvement of the on-site CV deployment.

The Clemson University Connected Vehicle Testbed (CU-CVT) is a CV deployment site, which includes unique heterogeneous wireless communication, data analytics for real-time data streaming and data archival capabilities to empower the users with capabilities to operate multiple and diverse CV applications simultaneously. This paper discusses the case study of the deployment of the CU-CVT with the goal of sharing the lessons learned from the CU-CVT deployment, which can be leveraged by other public agencies, and academic or research institutions. This motivation of this study is to promote early large-scale CV deployments by disseminating our findings in the deployment and operation of unique CU-CVT features.

The remainder of this paper is structured as follows. Section 2 describes existing connected vehicle testbeds and their limitations. Section 3 presents the CU-CVT and unique features of this testbed. Lessons learned from this testbed are presented in section 4. Finally, section 5 provides a concluding discussion.

**2.0 EXISTING CONNECTED VEHICLE TESTBEDS**

The USDOT leads the pilot deployment effort for Connected Vehicle Testbeds in the United States through the ITS Strategic Plan *(6)*. Southeast Michigan was the site of the first Connected Vehicle Testbed, which is sponsored by the USDOT, and is available to DSRC product developers and CV application developers to test any CV applications. Testbed development began in 2009, and the project was completed in August 2017. Subsequently, the USDOT extended support for the sharing of CV related data to develop a common technical platform for affiliated partner CV testbeds, such as the California and New York testbeds *(7)*. Key CV technologies are Signal Phase and Timing (SPaT), Geometric Intersection Description (GID) data broadcast, CV devices (i.e., Vehicle Awareness Devices (VADs), Aftermarket Safety Devices (ASDs), and Road Side Units (RSUs), and the USDOT preliminary Security Credentials Management System (SCMS) *(8)*. The two ongoing projects are the "5.9 GHz Dedicated Short-Range Communication Vehicle-based Road and Weather Condition Application: Phase 2" and "Basic Infrastructure Message Development and Standards Support". The purpose of the first project is to develop and test the 'weather condition information' transmission from CVs (i.e., public agency vehicles with connectivity used as CVs) to a Central Server via RSUs using 5.9 GHz DSRC communication. Public agency maintenance personnel use this stored information for future maintenance purposes.

In 2012, the University of Michigan Transportation Research Institute (UMTRI) and the USDOT started a new project to reduce roadway crashes using CV technology (known as Safety Pilot Model Deployment (SPMD) *(5)*. The total investment for this three year SPMD project (2015-2018) was $30 million, which incorporated over 2800 vehicles and 73 lane-miles of northwest Ann Arbor, Michigan, that was extended to the 27 sq. mile city of Ann Arbor with 45 street locations and 12 freeway sites. Every year 1500 vehicles are to be added in this deployment with the goal of having 5,000 vehicles on the road by 2018. The main goals of this testbed are to ensure that i) UMTRI and its partners will operate, maintain and upgrade this testbed from 2015 to 2018, ii) the Ann Arbor Connected Vehicle Test Environment will transition from research status to operational deployment, and iii) the UMTRI and its partners transition from a government-funded project to one that is self-sustaining. It will become the world's largest CV and infrastructure deployment and is called the Ann Arbor Connected Vehicle Test Environment (AACVTE).

The USDOT awarded a total of $45 million to sites in New York City (NYC), New York; Wyoming and Tampa, Florida in September, 2016 to initiate design, deployment and validate the Connected Vehicle (CV) Pilot Deployment Program *(9)*. The primary focus of the New York City pilot deployment is to improve vehicle and pedestrian safety *(10, 11)*. This pilot deployment includes 280 RSUs and 10,000 vehicles (cabs and buses), which will be used to evaluate CV safety challenges and benefits in an urban environment. Furthermore, they will evaluate pedestrian safety applications using 100 pedestrian-DSRC units. Multiple locations on Interstate 80 were chosen as the Wyoming CV pilot deployment site in another project *(12)*. This project deploys DSRC-OBU in 400 CVs (mainly snowplows and heavy vehicles) and provides safety alerts and travel guidance to the commercial CVs. In Florida, the Tampa-Hillsborough Expressway Authority (THEA) pilot deployment plans to implement a variety of V2V and V2I applications *(13)*. The focus of implementing these applications is to relieve congestion, reduce collisions, and prevent wrong way entry at the Selmon Reversible Express Lanes (REL) exit. Another purpose of the THEA pilot deployment involves using the CV technologies to improve pedestrian safety, accelerate bus operations and reduce conflicts between street cars, pedestrians and passenger cars at various locations. This deployment will also use DSRC as the data transmission medium for 1,600 cars, 10 buses, 10 trolleys, 500 pedestrians with smartphone applications, and approximately 40 RSUs along city streets.

The first public CV testbed in United States was developed in 2005 at California on El Camino Real (State Route 82) in Palo Alto, which consisted of 11 consecutive intersections between Stanford Ave at the north end and W Charleston Ave *(14)*. More than 50,000 vehicles travel this corridor between San Francisco and San Jose daily. The partners of this testbed are Caltrans, the Metropolitan Transportation Commission (MTC) and the California PATH program at UC Berkeley. In 2013, CV standards were implemented in the previously developed California CV testbed with the support of USDOT. The focus of this testbed is deploying Multi-Modal Intelligent Traffic Signal Systems (MMITSS) and Environmentally-Friendly Driving applications. The primary objective is to improve overall arterial network performance utilizing transit and freight signal priority, preemption for emergency vehicles, and pedestrian applications. Besides, Virginia two CV testbeds *(15), the* first, located in Southwest Virginia in Blacksburg Virginia at the Virginia Smart Road and the second, along Route 460 and Fairfax County in Northern Virginia along I-66 and on the parallel Routes 29 and 50, are the primary testing areas for the Connected Vehicle/Infrastructure University Transportation Center (CVI-UTC). Both were selected based on transportation system deficiencies, such as congestion, high crash rates and air quality non-attainment. Fifty roadside units were deployed along the roadways of both, which also includes different instrumented vehicles (e.g. instrumented automobiles, motorcycles, a motor coach and a semi-truck).

In existing CV testbeds, CVs are deployed to test CV safety applications using DSRC. Although DSRC is a very low latency wireless communication medium, it is limited in terms of communication range (~900ft (300m)) *(16)*. Thus, DSRC alone cannot support the breadth of CV applications (i.e., mobility and environmental applications) outlined by the connected vehicle reference implementation architecture (CVRIA) *(17)*. Other wireless technologies, such as LTE, WiMAX, and Wi-Fi can support DSRC communications to increase network availability and coverage area, and fulfil data rate requirements at peak period Thus, it is necessary to use heterogeneous wireless technology in which communication options can be chosen based on their range, data delivery delay, and throughput. Another limitation of existing testbeds is the missing data analytics platform for real-time data aggregation, processing and archiving for multiple CV applications running simultaneously. Designing a suitable data delivery system for CVs depends on the real-time data streaming from different devices without duplication and supplying the aggregated data depending on the application demand *(18)*. Appropriate computational resources are also necessary to support multiple and computationally expensive CV applications.

## 3.0 UNIQUE FEATURES OF CU-CVT TESTBED

The Clemson University - Connected Vehicle Testbed (CU-CVT) is developed in a hierarchical layered architecture. In a CV environment, different types of applications run at different levels (e.g. vehicle level, roadside infrastructure level, system level (i.e., backend server or cloud)) *(19)*. A layered architecture has proven effective in managing such distributed applications by decoupling the dependency and distributing the computation. The overview of the layered architecture in CU-CVT is shown in Figure 1. A CU-CVT layered architecture has a i) Mobile Edge device, ii) Fixed Edge Node, and iii) System Edge Node. The Mobile Edge Nodes will reside in Layer-1 of the architecture. Each Mobile Edge Node or each mobile entity (e.g. vehicles, pedestrians) is equipped with a DSRC, 3GPP/LTE Cellular or Wi-Fi enabled device that can collect and transmit location and mobility information. To support different type of applications with different levels of computation and memory requirements, DSRC devices are coupled with a computation unit capable of larger memory management, multiprocessor operations and user-friendly application development. The computation unit also permits the possible integration of other communication mediums, such as Bluetooth, Wi-Fi and LTE. The Fixed Edge acts as an entry point to the infrastructure with a very low latency with single hop DSRC connection with the Mobile Edge. Residing in Layer-2 of the system, the Fixed Edges communicate with components in Layer-1 and Layer-3. Fixed Edges can be extended to support a video camera and other sensing devices (i.e., weather sensors). Fixed Edges also contain a computational platform for the same reason as the Mobile Edge devices, along with a DSRC enabled device. Fixed devices are connected to Layer-3 with a high bandwidth, high throughput, low latency and highly reliable communication medium (e.g. fiber optics) to act as backhaul support.

The System Edge is situated at the top of the layered architecture, Layer-3. The task of the System Edge is to monitor the overall system to balance and optimize system operations and provide services to different nodes related to connectivity, data storage, application priority, dissemination of system wide messages and act as a secure portal for external agencies and research interests. One of the main responsibilities of the System Edge is to provide resource optimization and management of communication network resources in heterogeneous wireless networks. Using the Global Resource Controller (GRC) at the System Edge and Local Resource Controller (LRC) at the Mobile Edge, such requirements are fulfilled. Conceptually, the System Edge can be in a cloud or cluster of computers. We are using our own System Edge that supports the various services mentioned above.

Our CU-CVT system and CV applications is closely intertwined with Wireless Access in Vehicular Environments (WAVE) message sets stated in Society of Automotive Engineers (SAE) J2735 *(20)* and IEEE 1609.3 standards *(21)* even though we are not incorporating the WAVE Short Message Protocol (WSMP) standards. Our CU-CVT testbed provides users and developers the ability to operate a variety of applications and services that are interoperable with standard DSRC WAVE. Our applications comply with DSRC standards in terms of transmission channel, frequency and power *(22)*. Any other DSRC radios listening at standard DSRC channels would be able to operate with our applications without any problem. Our goal is to make the CU-CVT system highly interoperable with any other connected vehicle (CV) applications, which are compatible with WAVE multichannel standards (i.e., IEEE 1609.4) *(22)*. In addition, our testbed and the hierarchical architecture provides a solid framework for implementing security enforcing services. Our CU-CVT testbed and CV applications are designed to use the vendor specific security keys that are embedded into each licensed DSRC units. Our applications also provide methods to enable and disable such privacy enforcements. With the hierarchical structure of our testbed, implementation of Security Credentials Management Systems (SCMS) *(23)* or other forms of security and privacy methodologies can be easily implemented. In addition to the security and privacy platform available on DSRC networks, CU-CVT testbed has a virtual private network over the top of Clemson University campus network. The private network being used for the testbed provides an impenetrable backbone for data communication between Fixed Edge Nodes to a System Node. Collectively the private network and use of privacy keys make the test bed secured with application data privacy requirements satisfied.

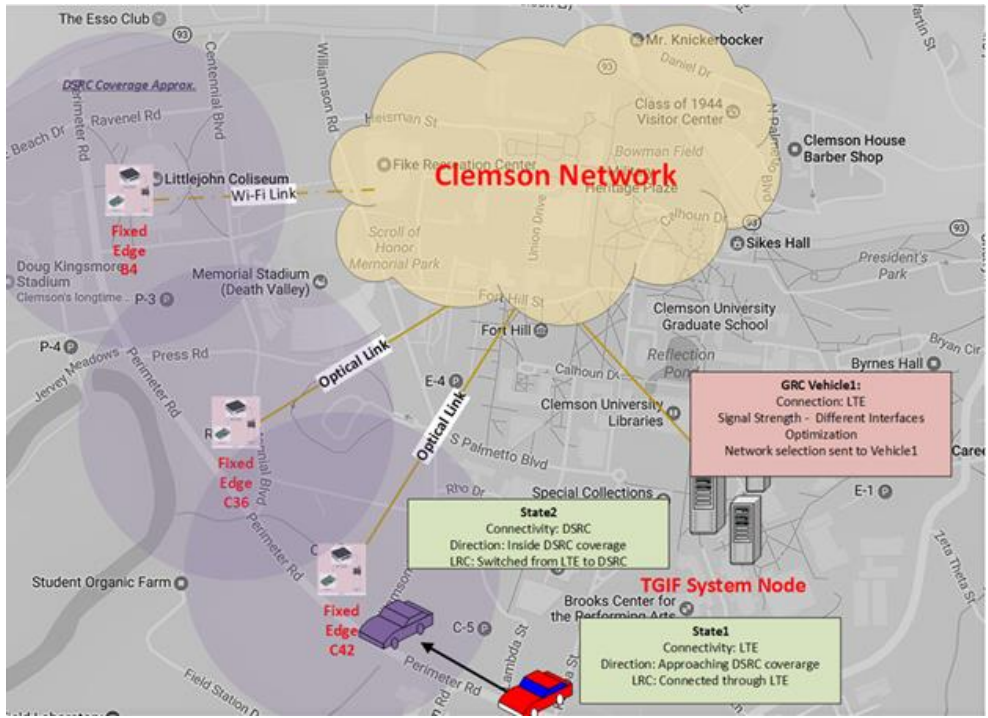

**FIGURE 1 Clemson University – Connected Vehicle Testbed (CU-CVT).**

## 4.0 LESSONS LEARNED: CU-CVT DEPLOYMENT

This section presents the lessons learned at different levels of the CU-CVT deployment: i) system design and prototyping; ii) system deployment and iii) system validation.

### 4.1 System Design and Prototyping

In the process of deploying the real world CV testbed, the first step involves defining the system architecture, selecting an appropriate communication medium for use among the CV components and software components, and evaluating and justify these selections by developing a prototype system. In this section, the lessons learned from the system design architecture and prototyping in a controlled environment are detailed.

*4.1.1 System Design Architecture*

The focus of this sub-section is to define the rationale for selecting a layered design architecture, heterogeneous wireless communication medium, data analytics using real-time data streaming and data archiving with big data management tools in the process of concept development of the CU-CVT.

**4.1.1.1 Layered Architecture** The key motivation for choosing the layered architecture of our testbed is to reduce data delivery delay for the services provided by a CVT testbed, reduce the bandwidth requirement and data loss rate to support a large number of vehicles and run multiple and diverse applications simultaneously *(19)*. DSRC provides a low latency, high throughput communication link from a Mobile Edge Node to a Fixed Edge Node and many applications, such as queue prediction, traffic data collection and video monitoring, which may need computation at the Fixed Edge rather than sending the data to a System Edge Node. Thus the overall effectiveness of the application is improved in terms of analyzing the data, which is a primary motivation for designing the layered architecture. Layered architecture can also support the integration of roadway sensors (e.g., video cameras, weather sensors, roadway traffic sensors) with the CVT deployment while reducing high bandwidth requirements. There are several advantages of this layered design architecture including proximity, intelligence, and scalability. *Proximity* of the edge provides the flexibility to communicate efficiently with the next immediate edge layer, and information distribution will be quicker than with a centralized system. The next immediate layer is defined as the closest edge layer in terms of the physical and logical sense. With the growing number of CVs, massive amounts of raw data will be collected and processed into usable formats for different CV applications. Thus, *intelligence* functions must be distributed smartly between different layers to reduce the computational time for computationally expensive applications. This layered architecture can be easily *scalable* by increasing the number of edges in different layers to meet the growing demand of computational resources.

**4.1.1.2 Heterogeneous Wireless Communication (Het-Net)** Traditional network management is complex, brittle, and error prone given the strong coupling of the data plane (for data forwarding) and control plane (for routing logic, packet management, and access control). Establishing the proper communication is a challenge in layered architecture. In order to ensure the reliability of data transfer for ITS applications, a combination of different wireless communication technologies such as –LTE, Wi-Fi, and WiMAX has been used *(17)*. Selection between different wireless communication options can be based on their feasibility, accessibility, and data delivery requirements of the CV applications *(24, 25, 26, 27, 28)*. Based on the CV application requirements, leveraging heterogeneous network technology for optimizing network resources provides a better solution to connectivity issues with CVT *(24)*.

**4.1.1.3 Data Analytics Using Real-Time Data Streaming** One of the primary challenges in layered architecture involves managing the data delivery system between different components. To mitigate this challenge, CU-CVT has adopted a publish-subscribe based data delivery system among the connected components *(18)*. A broker based system, which is a publish-subscribe based data delivery system, is a popular data delivery system in various areas, including social media, manufacturing, and e-commerce *(18)*. In general, in this system producers (e.g. sensors, RSUs) produce data, and consumers (e.g. traffic management centers, applications) consume data. Existing such frameworks include Kafka, RabbitMQ, ActiveMQ, WebsphereMQW, and MQTT *(18)*.

**4.1.1.4 Data Archiving Using Big Data Management Tools** Along with the data delivery system, a data storage system is also necessary. A database is a piece of software and storage area where we can store data that can be used for data analysis and reporting. In a CV environment, the different types of structured and unstructured data generated from different sources (e.g., on board units, RSUs, sensors) must be aggregated and stored at different edge levels of the layered architecture. Structured data can be stored in relational databases (Structured query languages (SQL)) and non-structured data is suitable for storage in NoSQL databases *(29)*. Oracle, MySQL, and SQL Servers are the most popular relational database systems used in industry, whereas MongoDB, CouchDB, and BigTable are the most popular NoSQL databases *(30, 31)*. Both have some advantages and disadvantages in terms of performance, usability, and scalability. In the section 4.3.3 of this paper, we describe our evaluation and selection of the most suitable data management tool for the CV environment.

*4.1.2 System Prototype Development in a Controlled Environment*

Prior to the real-world deployment of the system, it is necessary to develop a prototype of the application and system architecture in a controlled environment, especially in a controlled roadway for CV vehicles. Given that a focus of the CV applications is that of safety critical issues, deploying any application without such testing may create a hazardous situation. The first phase of application or research development involves testing the process in a controlled lab environment where the operation of the application, logic and errors are debugged and perfected. Therefore, we selected a particular corridor at Clemson University (i.e., the Student Organic Farm, as shown in Figure 2) as our real-world CV application test site. After the successful development and testing in a controlled environment, we deployed the application in the real world, which eventually involved CU-CVT.

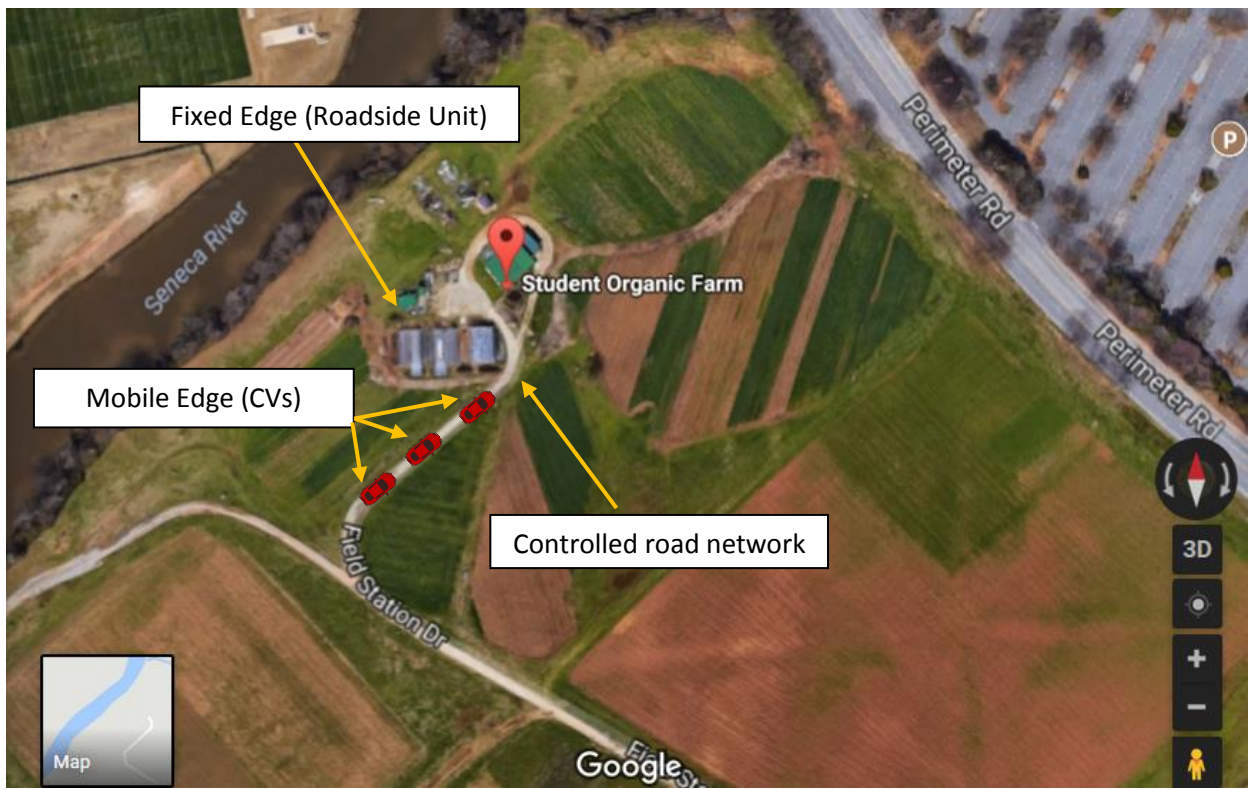

**FIGURE 2 Organic firm prototype deployment in a controlled environment.**

**4.2 System Deployment**
In this section, the lessons learned from the CU-CVT deployment are described in detail.

*4.2.1 Location Selection*

Identifying the optimal locations of physical devices is a critical aspect of field deployment. We have deployed our system on the main campus of Clemson University (Clemson, SC) as illustrated in Figure 3, with three Fixed Edge Nodes positioned along Perimeter Road. The location of the RSU depends on the surrounding environment, roadway geometry, opportunities for roadside mounting, and the purpose of the testbed. The specific range of functionality of the RSUs (owing to DSRC range limitation) requires placing them at intervals to ensure only a minimal range overlap. The blue line in Figure 3 shows the range from the field test.

The effect of RSU location on packet loss at two locations (C36 and C42) on the Perimeter Road deployment is shown in Figure 3. The RSU device used is the same for both RSU locations. One RSU shows higher coverage (Figure 3 (a)), while the other location (Figure 3 (b) shows less coverage because of the horizontal and vertical curve of the roadway. This is because of external factors such as interference and signal blockages. Thus, we need to select the location depending on the road geometry. The RSUs should be placed at locations with minimum interference from trees, houses and other signal blockages. For the CU-CVT deployment, we have chosen a corridor with multiple signals and used signalized intersections as the RSU locations. We consider an intersection as a suitable location which allows the use of RSU for CV safety, and mobility applications related to intersections (i.e., queue warning, stop sign gap assist). Simultaneous intersections have a certain spacing, allowing installation of one RSU at each intersection and covering the whole corridor. It also serves the purpose of covering multiple directions, which may be required for some CV applications (e.g., stop sign gap assist). Finally, the close proximity of the RSU to the signal controller provides easy access to the controller, which may also be valuable in V2I applications.

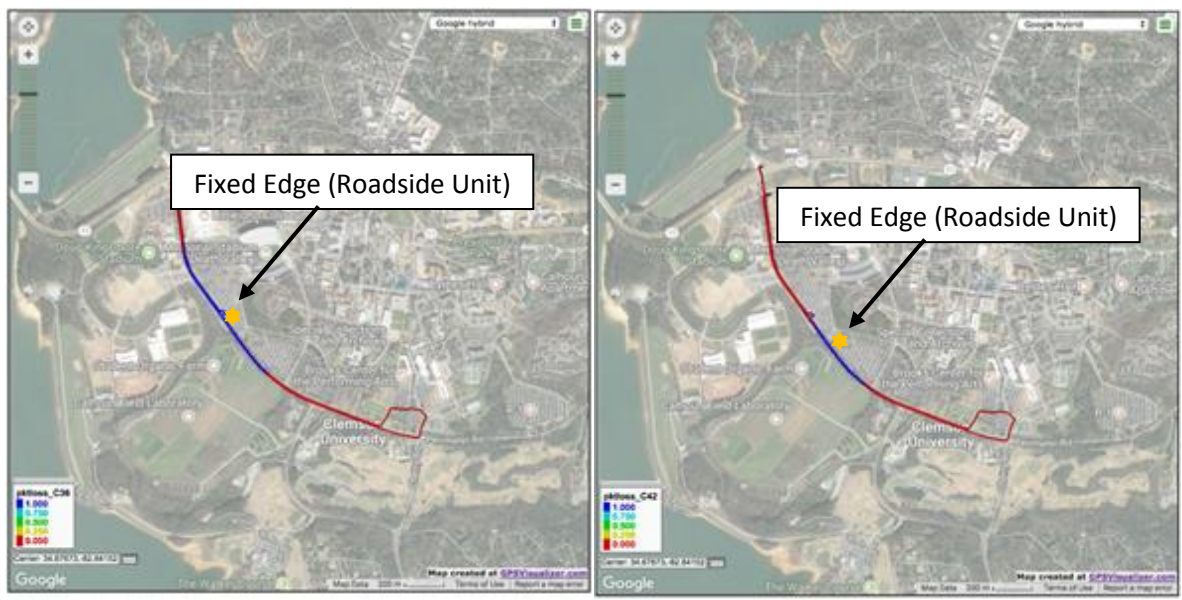

**FIGURE 3 (a) Packet loss at C 36 location.**  **FIGURE 3 (b) Packet loss at C 42 location.**

*4.2.2 System Compatibility and Integration*

One of the key issues for real-world deployment is system compatibility. From our deployment experiences, we found that different vendors have different ways of implementing applications, services, communication, and data streaming API. The main problem we experienced is that application development environment varies in different vendor devices. The testbed should be independent of vendor choice. As there is no specific standard for device configuration and it is the early phase of equipment production, one needs to write their own code to make it compatible with the platform. Another approach involves developing a generalized CVT distributed application services package, such as THings in a Fog (TGIF) *(19)* which overlays a linear and compliant wrapper to provide services for application developers.

*4.2.3 Heterogeneous Wireless Communication*

The application layer handoff is an elemental approach for managing heterogeneous networks, which as the name suggests involves the switching, or "handoff" from one network topology to another. In order to test the application layer handoff, the simple collision avoidance and traffic data collection applications were used. We need to modify the existing hardware or software of the OBU and RSU because of application layer handoff. The "link" and "IP" layers of the operating system control the hard handoff by prioritizing DSRC or Wi-Fi over LTE, which is detected by the application layer handoff. Two types of Het-Net technology were evaluated separately using the sample traffic data collection application, which is a simple V2I application that works with the Publish/Subscribe (Pub/Sub) scheme of data transfer from the Mobile Edge to the System Edge. Two types of Het-Net communication schemes were evaluated, a Wi-Fi plus LTE Het-Net and a DSRC plus LTE Het-Net. A hard-handoff is performed when the vehicle moves from the Wi-Fi coverage zone. A Transmission Control Protocol (TCP) and a User Datagram Protocol (UDP) handoff were used for the Wi-Fi plus LTE Het-Net.

*4.2.4 Data Analytics with Real-time Data Streaming*

The messaging service is equipped with a broker or the ability to publish/subscribe to a broker. For the CU-CVT deployment, we have used the MQTT (MQ Telemetry and Transport) broker *(32)*. MQTT is a lightweight open source publish/subscribe messaging transport protocol, and it is easy to implement. MQTT is superior than the Hypertext Transfer Protocol (HTTP) messaging service, because it provides better throughput, has lower power consumption, lower network bandwidth utilization and lower latency. The pub/Sub behavior makes the vehicular connectivity less onerous, as a client can directly communicate with the server. Pub/Sub decouples a client (message sender) from another client (message receiver) by making use of a unique set of strings for each type of message called Message Title. The MQTT broker filters all the received messages and distributes those messages accordingly. MQTT uses subject-based filtering of messages. CV applications are dynamic in nature, and most applications require efficient real-time data streaming and data distribution. Thus, data exchange in V2V and V2I communication is a major challenge. Moreover, CV applications operate in a constrained environment, where there are often limitations in bandwidth, computation and storage. Given the requirements of a lightweight broker based communication method, MQTT is a good choice for the field deployment of CV applications.

*4.2.5 Data Archiving with Big Data Management Tools*

Data storage is an essential part of CV applications. Many CV applications have to deal with big data, where there is a large volume of streaming data and a large variety of data (i.e., real time location, vehicle speed, position, acceleration). The choice of database for archiving data should be uniform across all layers of field deployment. For our CU-CVT deployment, we have used MongoDB as the uniform data storage platform for CV applications. As CV applications will need to deal with huge amounts of data, a non-relational database is a better choice than a relational one. NoSQL databases offer advantages over SQL databases in several aspects such as- scalability, growth and node addition. Among many NoSQL databases, MongoDB is chosen for the CU-CVT deployment because it has many advantages over other NoSQL databases. MongoDB is a popular open source NoSQL database tool for big data management. Being open source is a major advantage. MongoDB offers the query ability of a relational database while maintaining the features of a NoSQL database. Moreover, if the data is location based, MongoDB is handy because it has some built in spatial functions. MongoDB stores data using the very popular JSON format, which is compatible with all programming platforms. Different CV applications require different data storage requirements and different data structures. Therefore, we need a fast, scalable and flexible data storage platform that can store any type of data, of any size, at any time while the application is running. Moreover, different applications may operate on different platforms. We need a universal data storage platform that will perform effectively across all CV applications, and MongoDB is the solution.

*4.2.6 Inter-agency Cooperation*

The challenge of developing a testbed for field implementation requires collaboration with multiple stakeholders. For the CU-CVT deployment, significant collaboration was required between the researchers and the Clemson University Information Technology (IT) department. We often work closely with the IT department of the university, since we have used the campus Wi-Fi network (Eduroam) and optical fiber for backhaul support. To install a RSU at the testbed, we chose a light pole close to the traffic signal. We regularly communicated with the Clemson University campus physical planning department for the bucket truck to install the RSU at the Perimeter Road (where the CV-CVT is located) light pole. We selected the light pole so that there are optical fiber services from the pole to our backend resources. Without diligent support from the Clemson University campus physical planning and IT departments, the proper deployment of the testbed would have been impossible. The use of a state owned road requires permission from the state Department of Transportation (state DOT) to set up the roadside RSUs and secure any access to the state DOT infrastructure, specifically the DOT's communication network and traffic control devices and equipment.

## 4.3 System Validation in the CU-CVT

The lessons learned from system validation include validation experiences on DSRC system performance ascertainment, heterogeneous wireless network, data analytics with real time data streaming and system integration.

*4.3.1 DSRC System Performance Ascertainment*

The CV testbed is deployed on a geographically spread out network, including dynamic scenarios ranging from high speed-less congested scenarios to slow speed-highly congested scenarios. For these scenarios, it becomes very important to understand how the system performs to provide reliable communication between different Mobile Edge Nodes and Fixed Edge Nodes. The coverage range obtained for the three Fixed Edge Nodes in the CU-CVT is shown in Figure 4, with the Received Signal Strength Indicator (RSSI) value measured by the Fixed Edges. This coverage of the DSRC range provides an observation of how far the network can reach and allows for strategic planning for heterogeneous network management.

*Chowdhury, Rahman, Rayamajhi, Khan, Islam, Khan and Martin* 13

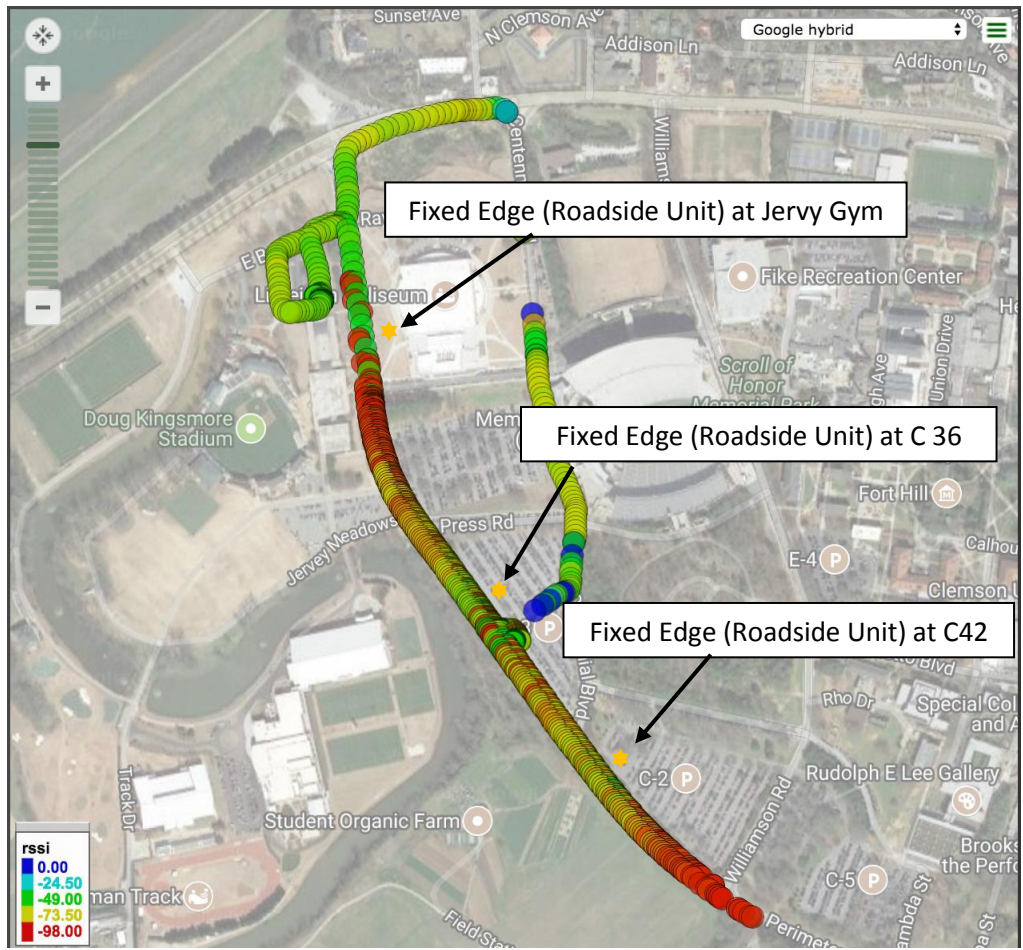

**FIGURE 4 DSRC network coverage of the CU-CVT testbed.**

*4.3.2 Heterogeneous Wireless Communication*

The average delay recorded between successive packet arrivals is used to calculate the handoff time. Vehicle speed data is continuously sent as data packets separately using TCP and UDP. Data packets are received and the delay is recorded. For each data packet, the 802.11 link must be activated and the vehicle must associate with the access point (RSU) to ensure a handoff delay. While the delay varies with successive data packets, a comparison of the longest delay between TCP and UDP determined UDP to have a lower delay of 25s compared to 28s for TCP. Let us consider a vehicle equipped with Het-Net technology that can send and receive data using DSRC and LTE simultaneously. RSU continuously broadcasts beacon-like messages within its coverage area limited by the range of DSRC communication. Prior to entering the coverage area of the RSU, the LTE communication is active in the vehicle. Since the vehicle is equipped with DSRC communication capabilities, it receives the beacon-like message when it enters the coverage area of the RSU and initiates the handoff process. The process executes a hard handoff from LTE to DSRC. The process also utilizes the backend infrastructure (i.e. a remote server) to store the data. The opposite handoff process (LTE to DSRC) is executed when the vehicle leaves the RSU coverage area and ceases receiving the beacon-like messages.

*4.3.3 Data Analytics with Real-time Data Streaming*

We collected raw data (i.e., time stamp, car ID, latitude, longitude and speed), known as Basic Safety Messages (BSMs) from each CV. The campus Fixed Edge used Wi-Fi to communicate with the backhaul server (System Edge). Vehicles equipped with CV On-Board equipment (i.e., Mobile Edge) could thus collect real-time data from three vehicles as those vehicles traveling within their DSRC range. All Mobile Edge device equipped vehicles send data to the RSU. Whenever one of the Mobile Edges is within the DSRC range of a Fixed Edge, the vehicle cluster connects to the RSU via DSRC. Optical fiber is used to connect with the System Edge layer when real-time communication is required. We host the database system using MongoDB to store data at the Fixed Edge and at the System Edge. After collecting raw data from the DSRC OBU, the DSRC RSU will publish raw data to the MQTT broker system, and later the raw data processing unit consumes data from the MQTT broker. The MQTT broker, the MongoDB database, and other computational processes reside in an edge-computing device. After processing the raw data, the processed data is stored at the MongoDB database in the Fixed Edge node device. CV applications can use the processed data from the Fixed Edge device. Data will be stored at the Fixed Edge device for a short period of time. Data can also be stored for longer periods in the MongoDB data storage at the System Edge.

*4.3.4 System Integration and Validation: A Case Study with Collision Avoidance and Queue Detection Application*

To verify the accurate integration of the hardware and software components with reliable communication systems to ensure the requisite safety and mobility CV benefits, the overall system integration was validated by testing the collision avoidance and queue detection applications.

**4.3.4.1 Collision Avoidance Application** From the field tests with the CU-CVT, we recorded the time required to deliver emergency safety messages from a vehicle that stopped suddenly (CV-1 or Mobile Edge 1) to the immediate following vehicle (CV-2 or Mobile Edge 2) using DSRC. We used three different vehicle speeds (i.e., 20 mph, 35 mph and 50 mph) for the field tests. Table 1 provides the average safety distance between CV-1 and CV-2 in three test runs with three different speeds (i.e., 20 mph, 35mph, and 50 mph). We also calculated the minimum safety distance requirement, for CV-2 (Table 1, Column 3) when CV-2 applied the brake after receiving the first warning message, with a deceleration rate of 11.2 ft/s$^2$ *(33)*. As shown in Table 1, the minimum safety distance required is less than the average safety distance available between CV-1 and CV-2 at all three vehicle speeds. In addition, the observed average message delivery latency is less than the latency requirements for CV safety applications *(34)*. The receiving time of the safety message in the vehicles downstream depends on several factors such as, vehicle movement patterns, vehicle speeds and any interference or noise in the environment. Thus, the distance to the upstream vehicle at the time of receiving the safety message varied for different tests. Simultaneously, the exchange of safety messages between CV 1 and CV 2 served as a prompt for the following three events:

1. Using LTE, the Het-Net system in CU-CVT delivered the collision-warning message from CV-1 and CV 2 to another vehicle (CV-3) which was beyond the DSRC range;
2. CV-1 sent BSM to the nearest Fixed Edge via DSRC as CV-1 was within its DSRC range;
3. The Fixed Edge sent the BSMs to the System Edge layer using the Wi-Fi communication option available in CU-CVT.

The LTE communication supported vehicle (CV-3), which was outside the DSRC range of CV-1, received the warning message with a longer delay with LTE, which is above the required latency for CV safety applications but within the latency requirement to receive possible upstream incident warning. This discretion in latency proves the suitability of Het-Net for CV applications enabling multiple wireless networking options based on the latency requirements of the CV applications. We hosted a database system using MongoDB to store data at the Fixed Edge and at the System Edge. After collecting BSMs from the DSRC enabled CV 1, the DSRC enabled RSU at the Fixed Edge will publish BSM to the MQTT broker system, and MongoDB archives BSMs from the MQTT broker. BSMs are also archived in the MongoDB database at the System Edge for future use. In these CU-CVT experiments, we validated the optimal utilization of available edge nodes (i.e., the Fixed Edge and System Edge) at different layers and with different communication options depending on the CV applications' communication and data analytics requirements.

**TABLE 1 Collision Avoidance Results for CV-2**

| Vehicle speed (mph) | Safety Distance | | Message Delivery Latency | | |
|---|---|---|---|---|---|
| | Average safety distance of three test runs between CV-1 and CV-2 (when 1st warning message packet received) (ft) | Minimum safety distance requirement (Considering 11.2 ft/s$^2$ deceleration rate) (ft) | DSRC Average Message Delivery latency (ms) | LTE Average Message Delivery latency (ms) | Latency Requirements for CV safety applications (ms) (34) |
| 20 | 305 | 38 | 88 | 2590 | 200 |
| 35 | 279 | 118 | 102 | 2810 | |
| 50 | 313 | 240 | 125 | 3000 | |

**4.3.4.2 Traffic Queue Detection Application** A queue ahead of a traveling vehicle is one of the major causes of rear-end collisions that can also disrupt traffic throughput by introducing shockwaves into the upstream traffic flow *(35)*. Daily recurring congestion, work zones, incidents and/or weather conditions are also causatives of traffic queues *(35)*. We can reduce rear-end collisions and the impact of resulting traffic flow shockwaves significantly through the rapid detection of a traffic queue, the prompt formulation of an appropriate response message for approaching vehicles, and the dissemination of queue information to the approaching vehicles in real-time. A threshold-based queue detection algorithm was developed by USDOT using V2V and V2I communication utilizing DSRC *(35)*. Typically, the threshold-based algorithm uses average speed of the vehicle (i.e., less than 5 mph) as well as the average separation distance (i.e., less than 20ft) within DSRC communication range. Each CV in the DSRC range can broadcast a series of Basic Safety Messages (BSMs) messages every 1/10th of a second. We implemented the threshold-based queue detection algorithm and evaluated queue detection application accuracy in our CU-CVT testbed. The queue detection application was used to effectively showcase the usage of different features of our testbed for developing and deploying connected vehicle applications. We used both the average speed and average separation distance to detect a traffic queue in our testbed.

An application running in the Mobile Edge Node broadcast vehicular information (i.e., time stamp, car ID, latitude, longitude and speed) to a Fixed Edge Node within the communication range of the DSRC. Using a python script at Fixed Edge Node, we calculated the separation distance between two immediate CVs using the latitude and longitude of each vehicle. The speed and separation distance are also averaged of all three connected vehicles for each second. Using the average speed and average separation distance, we detected queue by running the developed threshold based algorithm at Fixed Edge Nodes when three CVs stopped at the traffic signal in front of Jervy Gym location. The accuracy of the threshold-based queue detection method is 81% for three test runs and we run the threshold-based algorithm 167 times (every one second) for traffic queue detection. To calculate the accuracy of the threshold-based algorithm, we compare the threshold based queue prediction algorithm with field collected video camera data, which represents the ground truth data. As the accuracy of the algorithm depends on average separation distance between vehicles, the reason of lower accuracy is a higher separation distance between the CVs higher than the threshold because of the presence of non-connected vehicles between CVs. The backhaul network consisting of Wi-Fi network is used to connect the Fixed Edge Nodes with the System Node when real-time communication is required. A Big Data software, MongoDB was used to store data at both the Fixed Edge and at the System Edge. In addition, we measured the data exchange delay for exchanging raw data and queue detection information between the Edge nodes. A summary of data exchange delay between different edge nodes is provided in Table 2. This effort was used to demonstrate the suitability of our testbed for deploying both safety as well as mobility CV applications.

**TABLE 2 Summary of Data Exchange Delay**

| Data Exchange Scenarios | Type of Communication | Data | Average Data Exchange Delay (milliseconds) |
|---|---|---|---|
| Mobile Edge (CV) – Fixed Edge | DSRC | Basic Safety Messages | 4 ms |
| System Edge - Fixed Edge | Wi-Fi | Queue detection information | 6 ms |

## 5.0 CONCLUSIONS

Our CU-CVT has several unique features, such as layered architecture, heterogeneous wireless communication medium, data analytics using real-time data streaming and data archiving with big data management tools. The layered architecture is validated through system design process. As the layered design architecture is easily scalable depending on the deployment area and penetration level of CVs, the CU-CVT design is applicable for future large scale deployment. In addition, our developed heterogeneous wireless technique can be easily implemented in many other CV testbeds. The selection between different wireless communication options (Wi-Fi, WiMAX, LTE) is based on the feasibility, accessibility, and data delivery requirements of the application. The use of this heterogeneous network technology also makes the expansion of traffic data collection for traffic management possible in terms of the data collection coverage area in a CV environment. The testbed and supporting software Application Programming Interface (API)'s are designed to make the application developed for the testbed independent of both different network technologies and networking devices. The testbed also easily reconfigurable for forthcoming networking technologies like 5G-LTE. Our goal has been to develop applications and systems that can be easily integrated to any platform and systems, and that support any new networking technologies. The motivation has been to develop a CV connectivity strategy where the applications can be developed independently of networking and other CV device attributes. These features make the CU-CVT testbed and application development platform reach longer life cycle in terms of future CVT research, transferability of our work and supporting current and future CVT technology development.

Lessons learned were identified at three different levels: i) the development of design architecture and prototyping in a controlled environment, ii) the deployment of the CU-CVT testbed, and iii) the validation of the deployment with a CV application experiment. While all wireless communication technologies, such as DSRC, LTE, WiMAX, and Wi-Fi are easily accessible by CVs, DSRC is usually preferred for safety applications due to its very low latency. We evaluated the performance of heterogeneous wireless technologies, Wi-Fi, DSRC and LTE for the collision avoidance application and traffic data collection and archiving purposes. The field tests found that the heterogeneous wireless network can support multiple applications simultaneously, including safety and mobility applications. For data analytics in real-time data streaming, data aggregation and processing should be done based on on-demand data requests. Brokering systems can support data distribution for real-time data analytics by reducing data redundancy. In addition, real-time data archiving is required at the System Edge level for mobility and safety applications using big data management tools. The experimental results for the identified unique features of the CU CVT testbed would advance future connected vehicle research and serve as reference to agencies for the deployment of CV infrastructure in the real world. The CU-CVT was found to be an invaluable resource for evaluating CV technologies and applications.

## ACKNOWLEDGEMENTS

The authors wish to acknowledge the editorial assistance of Mr. Godfrey Kimball in the preparation of this draft.

This material is based upon work supported by the USDOT Connected Multimodal Mobility University Transportation Center ($C^2M^2$) (Tier 1 University Transpiration Center) headquartered at Clemson University, Clemson, South Carolina, USA and the National Science Foundation (NSF) under U.S. Ignite Grant #1531127. Any opinions, findings, and conclusions or recommendations expressed in this material are those of the author(s) and do not necessarily reflect the views of the USDOT Center for Connected Multimodal Mobility ($C^2M^2$), NSF and the U.S. Government assumes no liability for the contents or use thereof.

*Chowdhury, Rahman, Rayamajhi, Khan, Islam, Khan and Martin*